\documentclass[aps,pra,superscriptaddress,reprint,floatfix,longbibliography,showpacs]{revtex4-1}
\usepackage[english]{babel}
\usepackage{amsmath}
\usepackage{amssymb}
\usepackage{graphicx}
\usepackage{upgreek}
\usepackage{epstopdf}
\usepackage{color}
\usepackage{dcolumn}
\usepackage{multirow}
\usepackage[note-name=, use-sort-key = false]{notes2bib}
\usepackage[normalem]{ulem}

\usepackage{stackengine}

\usepackage[colorlinks,bookmarks=false,citecolor=darkblue,linkcolor=red,urlcolor=blue]{hyperref}
\definecolor{darkred}{rgb}{0.7,0.0,0.0}
\definecolor{darkblue}{rgb}{0,0.02,0.45}

\usepackage{xr}
\def\cm{cm$^{-1}$}

\begin{document}
\title{Charge dynamics in the Weyl semimetals NbIrTe$_4$ and TaIrTe$_4$ under pressure: Signatures of an electronic phase transition}

\author{M. Lamp}
\affiliation{Experimentalphysik II, Institute of Physics, University of Augsburg, 86159 Augsburg, Germany}
\author{J. Ebad-Allah}
\affiliation{Experimentalphysik II, Institute of Physics, University of Augsburg, 86159 Augsburg, Germany}
\affiliation{Department of Physics, Tanta University, 31527 Tanta, Egypt}
\author{A. Chmeruk}
\affiliation{Theoretische Physik III, Institute of Physics, University of Augsburg, 86159 Augsburg, Germany and Augsburg Center for Innovative Technologies (ACIT), University of Augsburg, 86159 Augsburg, Germany}
\author{N. Bura}
\affiliation{Experimentalphysik II, Institute of Physics, University of Augsburg, 86159 Augsburg, Germany}
\author{R. Sch\"onemann}
\affiliation{National High Magnetic Field Laboratory, Florida State University, Tallahassee, Florida 32306, USA}
\author{L. Balicas}
\affiliation{Department of Physics and Astronomy, Baylor University, Waco TX, 76798-7316, USA}
\author{S. H. Lee}
\affiliation{2D Crystal Consortium, Materials Research Institute, Pennsylvania State University, University Park, PA 16802, USA}
\affiliation{Department of Physics, Pennsylvania State University, University Park, Pennsylvania 16802, USA}
\author{Z. Q. Mao}
\affiliation{2D Crystal Consortium, Materials Research Institute, Pennsylvania State University, University Park, PA 16802, USA}
\affiliation{Department of Physics, Pennsylvania State University, University Park, Pennsylvania 16802, USA}
\affiliation{Department of Materials Science and Engineering, Pennsylvania State University, University Park, Pennsylvania 16802, USA}
\author{L. Chioncel}
\affiliation{Theoretische Physik III, Institute of Physics, University of Augsburg, 86159 Augsburg, Germany and Augsburg Center for Innovative Technologies (ACIT), University of Augsburg, 86159 Augsburg, Germany}
\author{C. A. Kuntscher}
\email{christine.kuntscher@physik.uni-augsburg.de}
\affiliation{Experimentalphysik II, Institute of Physics, University of Augsburg, 86159 Augsburg, Germany}

\begin{abstract}
A high-pressure investigation of the Weyl semimetals NbIrTe$_4$ and TaIrTe$_4$ is presented, using infrared spectroscopy supplemented by density functional theory calculations.
The experimental optical conductivity spectra as a function of pressure suggest the occurrence of a pressure-induced phase transition at a critical pressure $P_\text{c}=7\text{--}8$\,GPa. This transition is most likely electronic in nature, as Raman scattering measurements provide no evidence of a significant structural phase transition.
Above $P_\text{c}$ a significant redistribution of spectral weight occurs in the optical conductivity spectrum for both materials.
A Drude-Lorentz analysis of the optical data indicates a sharp reduction in the free carrier concentration at $P_\text{c}$, concomitant with the appearance of a low-energy phonon, which was initially screened by free charge carriers.
A predominantly electronic origin of the phase transition is supported by the calculated electronic band structure, Fermi surface, and interband optical conductivity as a function of pressure.
Our findings provide collective evidence for a pressure-induced, most likely electronic phase transition in both van der Waals materials at $P_\text{c}=7\text{--}8$\,GPa, highlighting the tunability of their electronic band structure by hydrostatic pressure.
\end{abstract}

\maketitle

\section{Introduction}
Weyl and Dirac semimetals are three-dimensional phases of matter hosting topologically protected, gapless electronic excitations~\cite{Armitage.2018, Yan.2017, Hasan.2017, Young.2012}.
A Dirac node can be viewed as two degenerate Weyl nodes of opposite chirality, which split into Weyl nodes when either inversion or time-reversal symmetry is broken~\cite{Armitage.2018}.
In these materials, low-energy quasiparticles can behave as Weyl fermions, emerging at so-called Weyl points where two linearly dispersing electronic bands cross. This phenomenon, long sought in particle physics, was first experimentally confirmed in the condensed matter system TaAs~\cite{Lv.2015}. A distinct class, the type-II Weyl semimetal phase was later proposed, where the Weyl cones are strongly tilted and the Weyl points appear at the contact between electron and hole pockets, a scenario permitted by the absence of Lorentz symmetry in crystals~\cite{Soluyanov.2015}. The material WTe$_2$ was the first candidate identified to host this new type of fermion, a prediction supported by unique transport signatures such as an anisotropic negative longitudinal magnetoresistance~\cite{Soluyanov.2015, Li.2017-WTe2}.
The ternary transition-metal iridium tellurides TaIrTe$_4$ and NbIrTe$_4$ have also been proposed as type-II Weyl semimetals.
In TaIrTe$_4$ and NbIrTe$_4$, angle-resolved photoemission spectroscopy reported surface Fermi-arc states, which are a direct indication of the presence of Weyl points~\cite{Zhou.2018, Ekahana.2020}.
First-principles calculations initially suggested that TaIrTe$_4$ constitutes an ideal minimal example, hosting only four type-II Weyl points - the lowest number allowed by symmetry in a time-reversal invariant, inversion-breaking system~\cite{Koepernik.2016}. The related compound, NbIrTe$_4$, was predicted to host a more complex configuration with sixteen Weyl points in the presence of spin-orbit coupling~\cite{Li.2017}. However, a subsequent combined theoretical and experimental study of TaIrTe$_4$ challenged its status as a minimal Weyl semimetal, revealing a more intricate electronic structure with a total of twelve Weyl points and a pair of nodal lines~\cite{Zhou.2018}.

\begin{figure*}[t]
\includegraphics[width=0.8\textwidth]{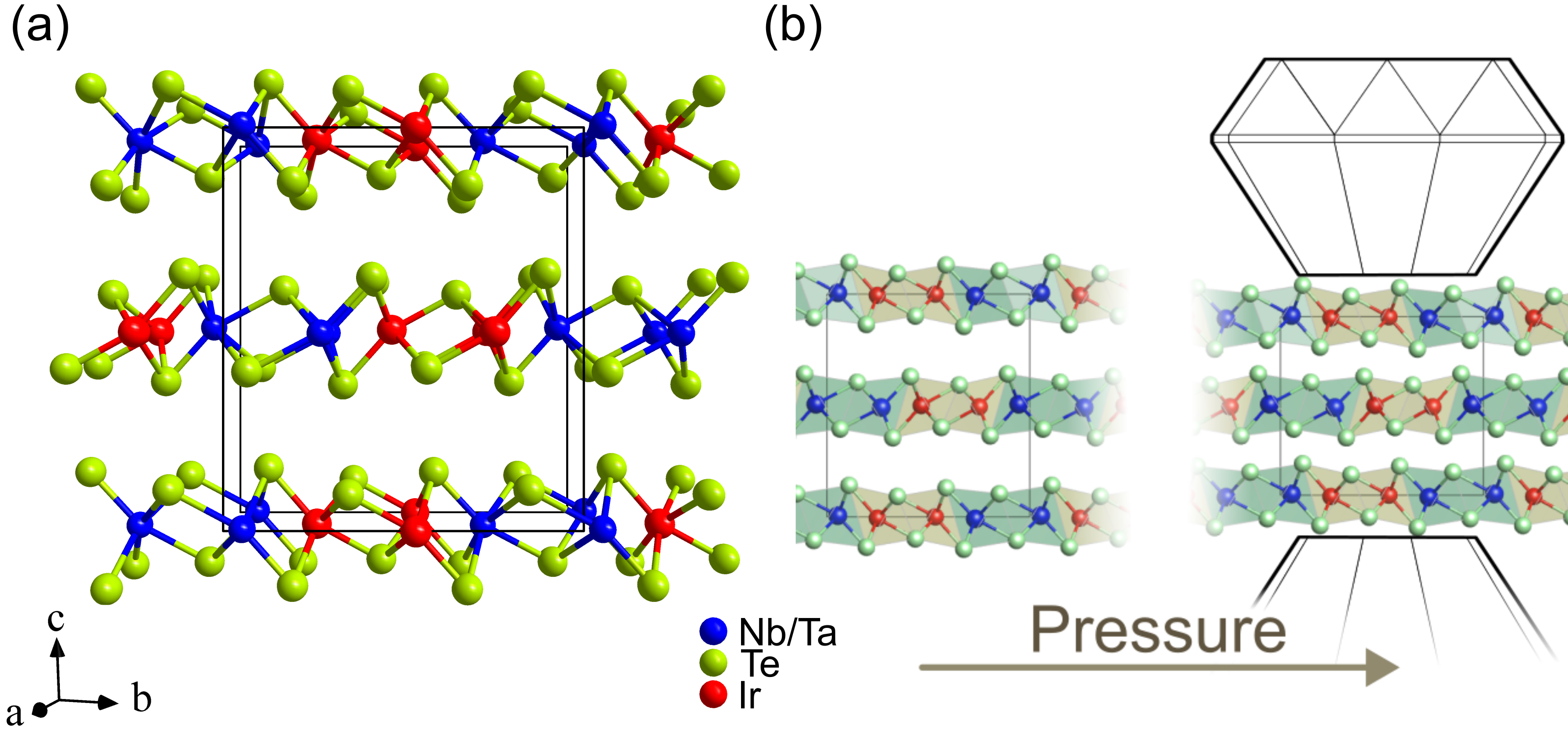}
\caption{Crystal structure and high-pressure compression scheme of NbIrTe$_4$ and TaIrTe$_4$. (a) Projection of the layered orthorhombic unit cell in the $bc$-plane. The atoms are color-coded: Nb/Ta (blue), Ir (red), and Te (green). The solid black rectangle indicates the unit cell boundary. (b) Schematic illustration of the high-pressure evolution. The application of hydrostatic pressure via a diamond anvil cell primarily compresses the soft van der Waals gap, leading to a significant decrease in the interlayer distance along the $c$ axis.}\label{fig:crystalstructure}
\end{figure*}

At ambient pressure, both NbIrTe$_4$ and TaIrTe$_4$ crystallize in a layered, orthorhombic structure belonging to the non-centrosymmetric space group $Pmn2_1$~\cite{Liu.2018, Chen.2019}. The crystal structure [see Fig.~\ref{fig:crystalstructure}(a)] is similar to that of WTe$_2$ but with zigzag chains of Nb/Ta and Ir atoms along the crystal direction $a$.
The absence of an inversion center, a prerequisite for their topological properties, was unambiguously confirmed for TaIrTe$_4$ by polarized Raman spectroscopy, which detected phonon modes that would be forbidden in a centrosymmetric structure~\cite{Liu.2018}. 

High pressure has been shown to be an effective tool for tuning the electronic and structural properties of NbIrTe$_4$ and TaIrTe$_4$ compounds, leading to complex phase diagrams. In general, layered chalcogenide materials with weak van der Waals interaction between the layers are prone to pressure-induced structural and electronic phase transitions \cite{Krottmüller.2020, Ebad-Allah.2019,Krottenmueller.2020b,Koepf.2024,Ebad-Allah.2025}. Recently, the relevance for pressure-induced interlayer sliding in driving phase transitions in chalcogenides has been pointed out~\cite{Tian.2026,Chi.2014}.
Previous high-pressure studies on NbIrTe$_4$ revealed evidence for an electronic phase transition between 4 and 12\,GPa, characterized by a reconstruction of the Fermi surface and a change from multiband to hole-dominated conduction~\cite{Mu.2021, Jin.2021}.
High-pressure Raman measurements support the electronic nature of this transition, as no major structural change was observed, although a subtle lattice distortion at a lower pressure of $\sim$~3.5\,GPa~\cite{Mu.2021} could not be ruled out. Below $\sim$~13\,GPa an anisotropic compressibility of the crystal lattice of NbIrTe$_4$ has been reported~\cite{Jin.2021}: According to the fast decrease in c/a and c/b ratios under pressure, the lattice is more compressible along the $c$ direction due to the relatively weak van der Waals interlayer bonding, as illustrated in Fig.~\ref{fig:crystalstructure}(b). At much higher pressures, both compounds exhibit pressure-induced superconductivity: In TaIrTe$_4$, superconductivity occurs at around 24\,GPa~\cite{Cai.2019}, whereas for NbIrTe$_4$, the reported onset pressure varies significantly across the studies, ranging from 2\,GPa to 39\,GPa, likely due to sample-dependent structural responses~\cite{Jin.2021, Long.2021, Mu.2021}.

Despite this progress, several open questions remain as the previous studies have not revealed consistent results. The precise nature of the low-pressure electronic phase transition in NbIrTe$_4$ is not fully understood, particularly as it is not accompanied by a distinct anomaly in Raman spectra~\cite{Mu.2021}. Furthermore, it is unclear whether TaIrTe$_4$ undergoes a similar low-pressure transition, as existing studies have focused primarily on the emergence of superconductivity at much higher pressures~\cite{Cai.2019}. A comprehensive understanding of how pressure tunes the interlayer van der Waals coupling to drive these electronic changes, and direct spectroscopic evidence of the evolution of the topological features under pressure, is still lacking.

In this work, we study the charge dynamics of NbIrTe$_4$ and TaIrTe$_4$ single crystals under pressure by infrared spectroscopy over a broad frequency range. Our investigation is distinct from earlier ones, as we characterize both materials under the same experimental conditions (same spectroscopic setups, pressure transmitting medium etc). The experimental results are supplemented by density functional theory calculations of the pressure-dependent electronic band structure and optical conductivity. 
We observe clear signatures consistent with a reversible, most likely electronic phase transition in both materials at a critical pressure $P_c\sim7\text{--}8$\,GPa. These findings provide new insights into the pressure-tunable electronic landscape of this important class of topological semimetals.

\section{SAMPLE PREPARATION, EXPERIMENTAL AND
COMPUTATIONAL DETAILS}

Single crystals of NbIrTe$_4$ and TaIrTe$_4$ were grown using the Te-flux method~\cite{Canfield.1992,Schonemann.2019} and characterized by energy-dispersive x-ray spectroscopy, and x-ray diffraction.
Reflectivity measurements at room temperature were performed using an infrared microscope (Bruker Hyperion) equipped with a 15$\times$ Cassegrain objective and coupled to a Bruker Vertex 80v Fourier-transform infrared spectrometer.
These measurements were taken in a frequency range of 170 to 18\,000\,\cm, with a resolution of 2\,\cm\ in the far-infrared and a resolution of 4\,\cm\ in the mid-infrared to visible range.
Prior to the measurements, the samples were freshly cleaved to ensure optimal surface quality. Pressure was applied using an EasyLab diamond anvil cell (DAC).
This DAC has a culet size of 600$\,\upmu$m and we used a CuBe gasket. CsI was used as the pressure-transmitting medium to establish hydrostatic pressure conditions within the cell. The pressure inside the cell was determined using the ruby luminescence technique~\cite{Mao.1986, Syassen.2008}.

The reflectance at the diamond–sample interface in the DAC was calculated using the equation $R_\mathrm{s-d}=I_\mathrm{sample}/I_\mathrm{reference}$, where $I_\mathrm{sample}$ stands for the intensity reflected by the sample and $I_\mathrm{reference}$ stands for the intensity reflected by the CuBe gasket.
In the NIR and visible range, intensity $I_\mathrm{b}$ reflected from the lower diamond-air interface of the empty DAC was used as reference and the reflectance was calculated according to $R_\mathrm{s-d}= R_\mathrm{dia} \cdot I_\mathrm{sample}/I_\mathrm{b}$, where $R_\mathrm{dia} = 0.1667$ represents the pressure-independent reflectivity of diamond~\cite{Eremets.1992}. Possible time-dependent fluctuations of the experimental setup such as light source intensity were compensated by normalizing with the intensity reflected from the top diamond surface of the DAC.
We interpolated the measured spectra in the frequency range 1600–2650\,\cm, which is strongly affected by the multiphonon absorption of diamond~\cite{Mildren.2013}.
This interpolation, along with extrapolation to 0\,\cm, is based on a Drude-Lorentz fit.
The Drude-Lorentz model effectively combines the Drude description of free electron conductivity with the Lorentz model for atomic dipole oscillators.
Additionally, by using x-ray atomic scattering functions for higher-frequency extrapolation~\cite{Tanner.2015}, we can apply the Kramers–Kronig relations to calculate other optical functions, such as the complex optical conductivity 
$\sigma(\omega)=\sigma_1(\omega)+i\sigma_2(\omega)$
and the dielectric function, $\epsilon(\omega)=\epsilon_1(\omega)+i\epsilon_2(\omega)$.
The standard Kramers–Kronig relationship between reflectance and phase has been modified for taking into account the sample-diamond interface, $R_{\mathrm{s-d}}$~\cite{Pashkin.2006}.

\begin{figure*}[t]
\includegraphics[width=0.8\textwidth]{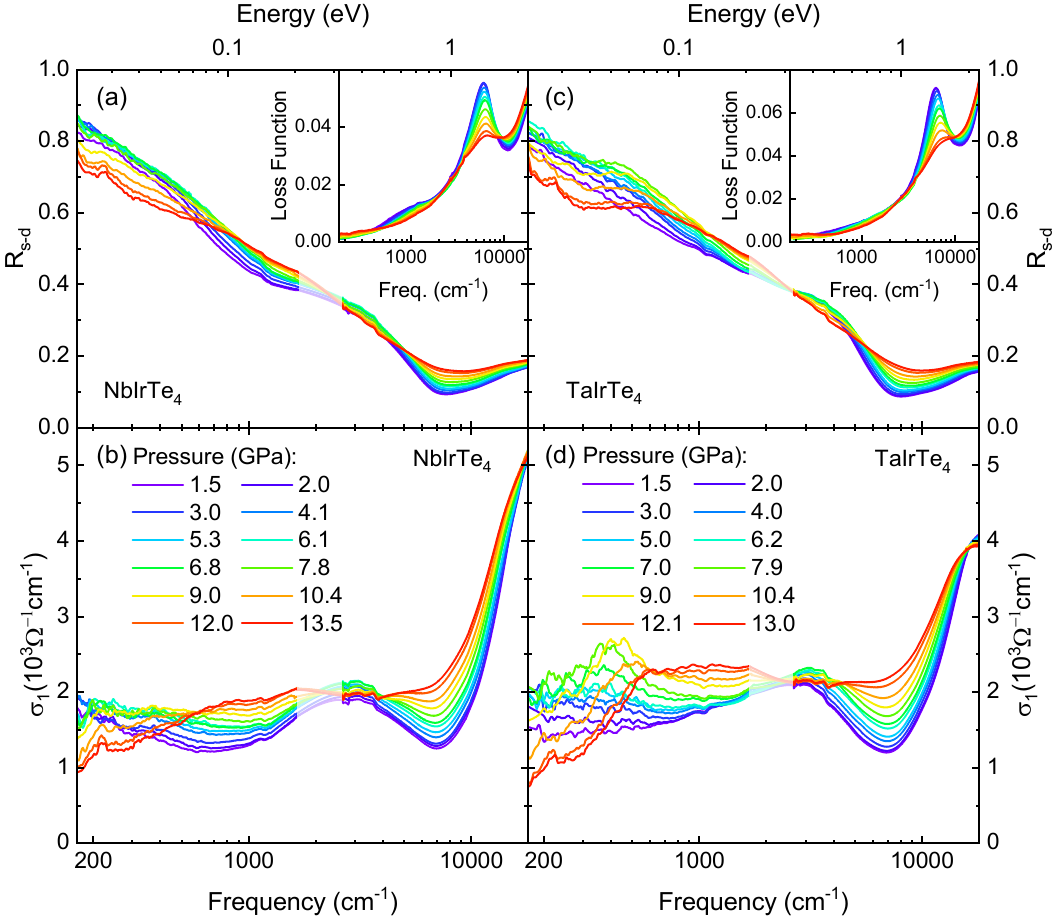}
\caption{Pressure-dependent (a) reflectivity ($R_{\mathrm{s-d}}$) and~(b) real part of the optical conductivity ($\sigma_1$) for NbIrTe$_4$. Pressure-dependent (c) reflectivity ($R_{\mathrm{s-d}}$) and~(d) real part of the optical conductivity ($\sigma_1$) for TaIrTe$_4$. The insets in (a) and~(c) show the corresponding loss functions as a function of pressure. The region affected by diamond absorption (1600–2650\,cm$^{-1}$) is highlighted in lighter colours.}\label{fig:infrared}
\end{figure*}

First-principles calculations were performed using the Vienna Ab initio Simulation Package (VASP)~\cite{Kresse.1996, Kresse.1996.2} based on Density Functional Theory (DFT)~\cite{Hohenberg.1964, Kohn.1965}. The projector-augmented wave (PAW)~\cite{Bloechl.1994} method  was used as implemented in the VASP code. We used the Perdew-Burke-Ernzerhof parametrization for solids (PBEsol) as exchange-correlation functional~\cite{Perdew.2008}.
The crystal structures of NbIrTe$_4$ and TaIrTe$_4$ were fully relaxed until the forces on each atom were less than $10^{-3}$ eV/Å. A plane-wave basis set with a kinetic energy cutoff of 390\,eV for NbIrTe$_4$ and 290\,eV for TaIrTe$_4$ was used. The Brillouin zone was sampled with a $12 \times 4 \times 3$ $\Gamma$ centered mesh.
The electronic self-consistency loop was converged to $10^{-8}$\,eV.
Gaussian smearing with a width of $\sigma = 0.1$\,eV was applied for all calculations.
The frequency-dependent optical conductivity was calculated using the Kubo-Greenwood formula~\cite{Pizzi.2020}. To obtain a converged spectrum, we first constructed a set of maximally localized Wannier functions (MLWFs) using the Wannier90 code~\cite{Marzari.1997, Mostofi.2014}. The initial projections for the Wannier functions were chosen as the d-orbitals for Nb/Ta and Ir, and the p-orbitals for Te.
From this Wannier representation, the electronic band structure, the Fermi surface, and the momentum operator matrix elements were interpolated from the coarse DFT grid onto a much denser 100$\times$50$\times$50 k-point mesh. Finally, the optical conductivity was evaluated for photon energies up to 2.2\,eV with a step of 0.005\,eV.
Test calculations including a van der Waals correction using the DFT-D3 method with Becke--Johnson damping function~\cite{Grimme.2011D3BJ} resulted in less accurate relaxed lattice parameters compared to plain PBEsol. The calculated optical conductivity also showed no qualitative difference, therefore we used PBEsol throughout.

\section{RESULTS}

\subsection{Infrared spectroscopy results}

\begin{figure*}[t]
\includegraphics[width=1\textwidth]{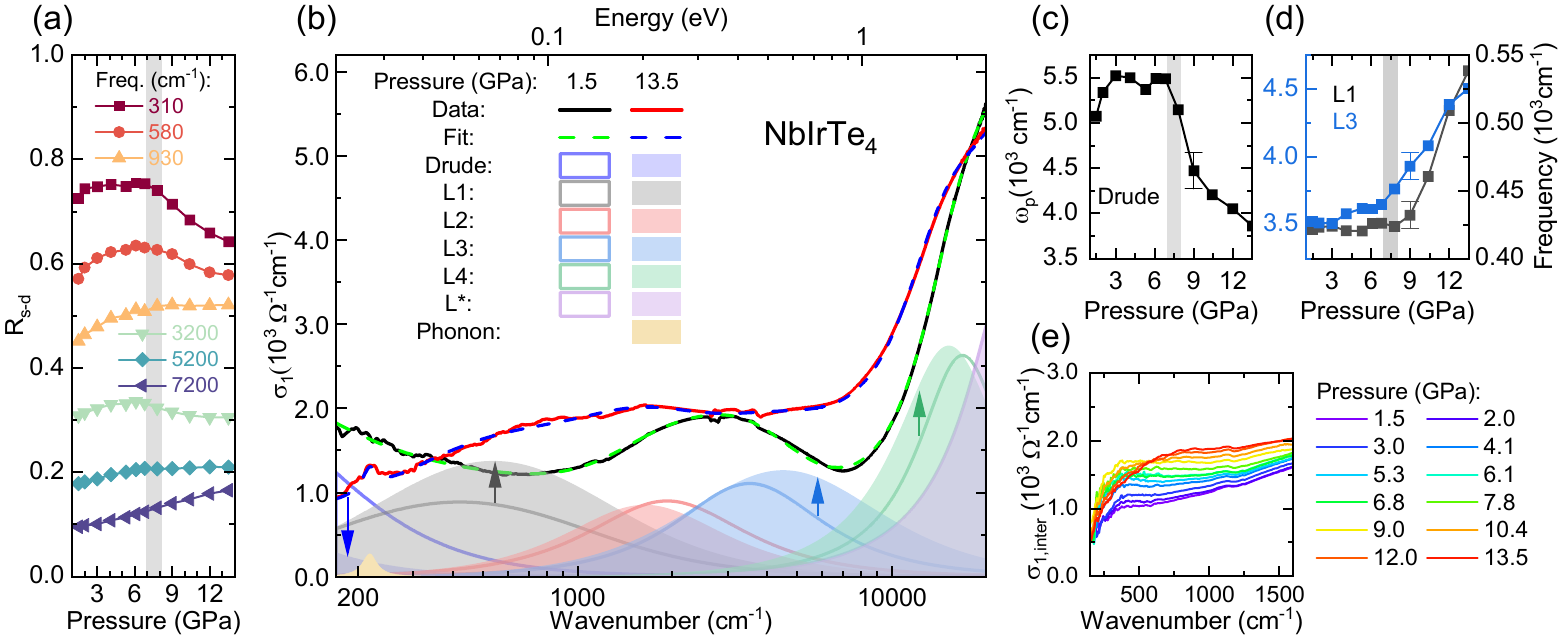}
\caption{Pressure-dependent optical properties and Drude-Lorentz modeling of $\text{NbIrTe}_4$. (a) Pressure evolution of the reflectivity $R_\text{s-d}$ at selected frequencies. (b) Optical conductivity spectra ($\sigma_1$) at 1.5\,GPa and 13.5\,GPa. The experimental data (solid lines) are fitted (dashed lines) using a Drude-Lorentz model. The shaded areas represent the contributions from the Drude term, phonon modes, and various Lorentz oscillators (L1–L4, L*) used in the fit. (c) Evolution of the Drude plasma frequency $\omega_\text{p}$ as a function of pressure. (d) Pressure dependence of the center frequencies of the L1 and L3 Lorentz oscillators. (e) Pressure evolution of the interband optical conductivity $\sigma_{1,\text{inter}}$ ($\sigma_1-\sigma_\text{Drude}$), obtained by subtracting the Drude contribution from the measured optical conductivity. The vertical gray bar across panels indicates the pressure region where anomalous behavior is observed. Representative error bars are displayed at 9 GPa in (c) and (d).}\label{fig:Nbparameters}
\end{figure*}

\begin{figure*}[t]
\includegraphics[width=1\textwidth]{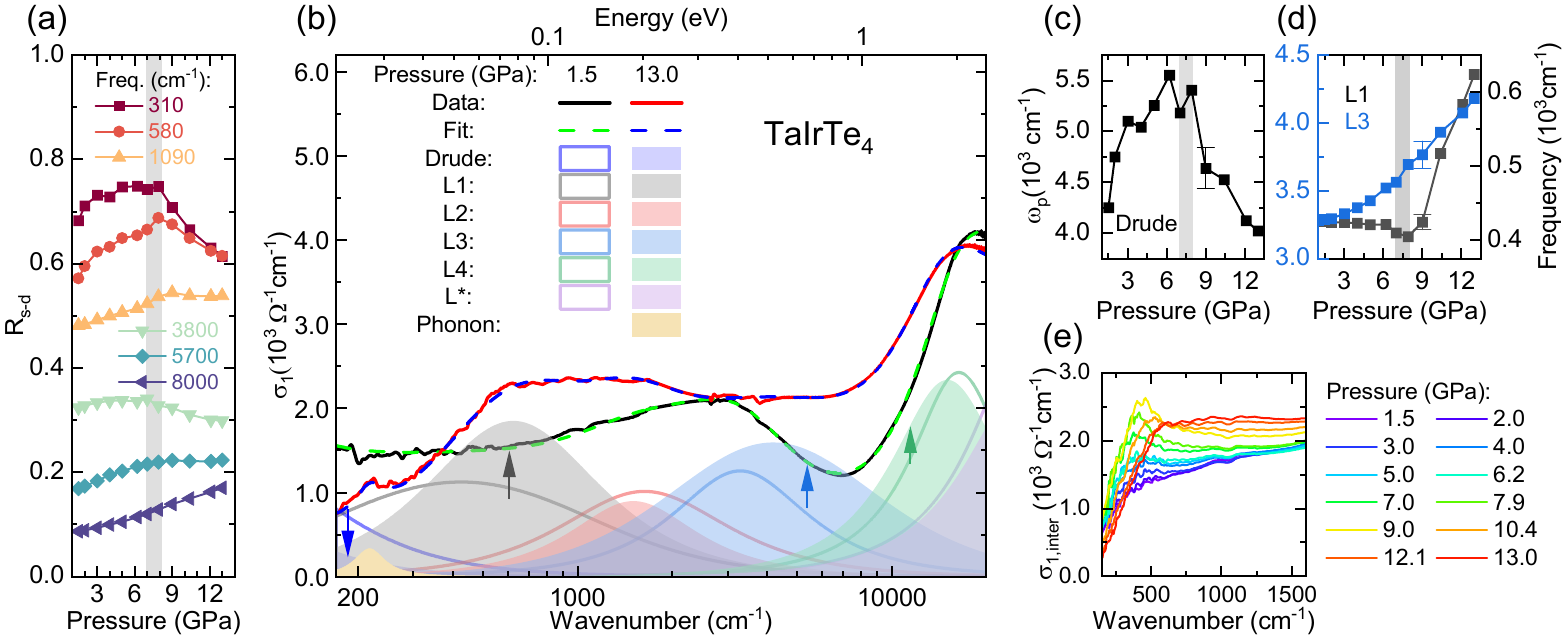}
\caption{Pressure-dependent optical properties and Drude-Lorentz modeling of $\text{TaIrTe}_4$. (a) Pressure evolution of the reflectivity $R_\text{s-d}$ at selected frequencies. (b) Optical conductivity spectra ($\sigma_1$) at 1.5\,GPa and 13.0\,GPa. The experimental data (solid lines) are fitted (dashed lines) using a Drude-Lorentz model. The shaded areas represent the contributions from the Drude term, phonon modes, and various Lorentz oscillators (L1–L4, L*) used in the fit. (c) Evolution of the Drude plasma frequency $\omega_\text{p}$ as a function of pressure. (d) Pressure dependence of the center frequencies of the L1 and L3 Lorentz oscillators. (e) Pressure evolution of the interband optical conductivity $\sigma_{1,\text{inter}}$ ($\sigma_1-\sigma_\text{Drude}$), obtained by subtracting the Drude contribution from the measured optical conductivity. The vertical gray bar across panels indicates the pressure region where anomalous behavior is observed. Representative error bars are displayed at 9 GPa in (c) and (d).
}\label{fig:Taparameters}
\end{figure*}

In Fig.~\ref{fig:infrared}(a) and (c) we display the pressure-dependent reflectivity, $R_{\mathrm{s-d}}$, for NbIrTe$_4$ and TaIrTe$_4$, respectively, measured at the sample–diamond interface across the 170--18\,000\,\cm range. The diamond multiphonon absorption region (1600--2650\,\cm) is highlighted in lighter colors, where it is bridged by a Drude-Lorentz fit. The lowest measured pressure was 1.5\,GPa, necessary to ensure a good sample-diamond interface, while the highest pressures reached were 13.5\,GPa for NbIrTe$_4$ and 13.0\,GPa for TaIrTe$_4$.
For the lowest pressure of 1.5\,GPa, both compounds show a high reflectivity $R_{\mathrm{s-d}}$ of 0.8 for low frequencies, indicating a metallic behavior. With increasing frequency, $R_{\mathrm{s-d}}$ continuously decreases to the level 0.4.
At around 6000\,\cm, we see a sharp decrease, which corresponds to the plasma edge.
In NbIrTe$_4$, a strong decrease in reflectivity can also be seen in the range above $\sim$~600\,\cm, which indicates a second plasma edge, whereas in TaIrTe$_4$ only one plasma edge is observed.
For both compounds we did not detect signatures of phonon modes in the $R_{\mathrm{s-d}}$ spectrum at 1.5\,GPa.

With increasing pressure, the reflectivity $R_{\mathrm{s-d}}$ of NbIrTe$_4$ for frequencies below 800\,\cm remains largely stable at a level of 0.8 up to the critical pressure $P_\text{c}=7\text{--}8$\,GPa, above which it decreases continuously to 0.7 at the low-frequency limit.
At around 3000\,\cm, the reflectivity shows a similar behavior, increasing up to a critical pressure $P_\text{c}$ and subsequently reducing. In contrast, at around 1800\,\cm and 7000\,\cm, the reflectivity increases continuously with pressure.
TaIrTe$_4$ shows a very similar behavior with an increase up to $P_\text{c}$, followed by a decrease in the range of below 1000\,\cm and between 2500\,\cm and 5000\,\cm.
At 9.0\,GPa in NbIrTe$_4$ and 10.4\,GPa in TaIrTe$_4$, a sharp feature emerges in the spectra, which we assign to a phonon mode near $\sim$220\,\cm. At lower pressure, Ref.\ \cite{Mardele.2020} reports this mode at a lower frequency, close to the low-frequency cutoff of our measurement range. In the present DAC measurements, this cutoff is limited by diffraction to about 170\,cm$^{-1}$ because of the small sample size.
Consequently, at low pressure the mode lies near the edge of our accessible spectral window and is difficult to resolve reliably.
In addition, this phonon can be screened at low pressure by the dominant free-carrier (Drude) contribution.
The pressure-induced reduction of the Drude contribution weakens this screening effect, allowing the previously obscured phonon to become a distinct and observable feature in the reflectivity spectrum.
An additional reason for the absence of this phonon in our low-pressure spectra may be the comparatively large Drude spectral weight of the present sample. For TaIrTe$_4$, our Drude-Lorentz analysis gives a plasma frequency of about 4500\,\cm ($\approx$0.55\,eV) at 1.5\,GPa and 300\,K, whereas \citeauthor{Mardele.2020} reported a smaller value of $\sim$0.1\,eV at ambient pressure and 5\,K~\cite{Mardele.2020}. Since the Drude spectral weight scales with $\omega_\mathrm{p}^2$, this corresponds to a roughly 30-fold larger Drude spectral weight and therefore a stronger metallic response in the present experiment. This stronger free-carrier response can enhance the screening of an infrared-active phonon and reduce its visibility in reflectivity. The difference in $\omega_\mathrm{p}$ may originate from a combination of the different experimental conditions (finite pressure and room temperature in the present DAC measurements versus ambient pressure and low temperature in Ref.\ \cite{Mardele.2020}) and sample-dependent variations in carrier density or effective mass, to which semimetals are particularly sensitive.

From the dielectric function $\epsilon(\omega)=\epsilon_1(\omega)+i\epsilon_2(\omega)$, as obtained by Kramers-Kronig analysis, the loss function, defined as $-\text{Im}(1/\epsilon)$, can be calculated. The loss function for NbIrTe$_4$, shown as an inset in Fig.~\ref{fig:infrared}(a), exhibits two distinct peaks corresponding to plasmon excitations.
A weaker plasmon peak is observed at approximately 800\,\cm, which vanishes completely at high pressures.
A second, more prominent peak is located at 6000\,\cm. This feature is also substantially suppressed with increasing pressure.
The loss function for TaIrTe$_4$ (inset in Fig.~\ref{fig:infrared}(c)) only shows the prominent peak at around 6000\,\cm which similarly decreases with pressure.

To better visualize the pressure-induced changes, Fig.~\ref{fig:Nbparameters}(a) and \ref{fig:Taparameters}(a) show the pressure dependence of the reflectivity $R_\mathrm{s-d}$ at selected frequencies for NbIrTe$_4$ and TaIrTe$_4$, respectively. In both materials, the behavior is non-monotonic and highlights the critical pressure $P_\text{c}$.
For NbIrTe$_4$, at lower frequencies (e.g., 310 and 580\,\cm), the reflectivity initially increases, reaching a maximum at the critical pressure $P_\text{c}$, before decreasing. A similar, though less pronounced, trend is observed at 3200\,\cm. At 930\,\cm and 5200\,\cm, the reflectivity is mostly flat above the critical pressure. For frequencies at 7200\,\cm and above, the reflectivity shows a continuous, monotonic increase with pressure.
TaIrTe$_4$ exhibits a very similar trend, as shown in Fig.~\ref{fig:Taparameters}(a), though at slightly different frequencies. These findings are consistent with a pressure-induced phase transition in both compounds, characterized by a common critical pressure where the reflectivity trends invert at low and mid-infrared frequencies.
Overall, the strong similarities in the pressure-dependent reflectivity spectra of both compounds suggest a common nature of the transition in NbIrTe$_4$ and TaIrTe$_4$.

Figures \ref{fig:infrared}(b) and (d) present the real part of the optical conductivity, $\sigma_1$, for NbIrTe$_4$ and TaIrTe$_4$ under pressure, calculated from the reflectivity spectrum $R_{\mathrm{s-d}}$ via Kramers-Kronig analysis.
For the lowest pressure of 1.5\,GPa, the optical conductivity of NbIrTe$_4$ [Fig.~\ref{fig:infrared}(b)] shows a slight decrease with increasing frequencies in the low-frequency range, with a minimum around 800\,cm$^{-1}$, followed by a broad absorption feature centered at approximately 3000\,cm$^{-1}$. A large absorption band then emerges starting from 10000\,cm$^{-1}$ and extending to higher energies. 
For TaIrTe$_4$ at 1.5\,GPa [Fig. \ref{fig:infrared}(d)], the optical conductivity at low frequencies is relatively constant, followed by a broad absorption band at around 3000\,cm$^{-1}$ and a pronounced absorption band starting at 10000\,cm$^{-1}$, very similar to the $\sigma_1$ spectrum of NbIrTe$_4$.

The optical conductivity spectrum of TaIrTe$_4$ at the lowest measured pressure (1.5\,GPa) is in good qualitative agreement with previous studies at ambient pressure by~\citeauthor{Mardele.2020}~\cite{Mardele.2020}, who investigated the optical response of TaIrTe$_4$ in the context of its nature as a type-II Weyl semimetal. In particular, below 40 meV they identified a nearly linear-in-frequency dependence of the interband optical conductivity at low temperatures.
It is important to note that the profile of the optical conductivity above 40\,meV was not discussed in detail in Ref.~\cite{Mardele.2020}, in particular the flat, plateau-like behavior, which we also observe in our data.
A linear-in-frequency increase in $\sigma_\text{1, inter}$ is also consistent with the presence of an energy-dispersive nodal line. In this case, the linear-in-frequency increase is followed by a plateau at higher energies~\cite{Shao.2019}.
Figures~\ref{fig:Nbparameters}(e) and \ref{fig:Taparameters}(e) show the interband optical conductivity $\sigma_{1,\text{inter}}$ ($\sigma_1$ without the Drude contribution $\sigma_\text{Drude}$ as obtained from the Drude-Lorentz model), which exhibits strong similarities to the expected features of a dispersive nodal line, particularly at higher pressure.
The plateau-like behavior starts in the frequency range at around 500\,\cm, and for higher pressure shifting to higher energies and extending to up to 1200\,\cm for NbIrTe$_4$ and up to the diamond absorption range for TaIrTe$_4$.

Below $P_\text{c}$, the optical conductivity of both NbIrTe$_4$ and TaIrTe$_4$ becomes relatively flat in the low frequency range with increasing pressure. 
Above this critical pressure, a significant reshuffling of spectral weight from low to high frequencies occurs in the $\sigma_1$ spectrum of NbIrTe$_4$: for frequencies below 800\,\cm, $\sigma_1$ abruptly drops, whereas it is enhanced in the 800–2000\,\cm range. A slight decrease is also observed in the feature around 3000\,\cm. TaIrTe$_4$ exhibits a comparable evolution, although with notable differences. As pressure increases, a distinct feature emerges at around 380\,\cm before the low-frequency conductivity is ultimately suppressed above the critical pressure. Furthermore, at the highest pressure of 13.0\,GPa, the conductivity shows a linear increase up to 600\,\cm, followed by a plateau-like behavior that extends to the diamond absorption region at 1700\,\cm. This pressure evolution is, in principle, also observed in NbIrTe$_4$, but the features are less clearly developed.

Further insights into the pressure-induced changes in NbIrTe$_4$ and TaIrTe$_4$ are gained from the extracted Drude-Lorentz model parameters, presented in Figs. \ref{fig:Nbparameters} and \ref{fig:Taparameters}, respectively. The Drude-Lorentz fits for the lowest and highest measured pressures are shown for NbIrTe$_4$ in Fig.~\ref{fig:Nbparameters}(b), and for TaIrTe$_4$ in Fig.~\ref{fig:Taparameters}(b).
One Drude term was sufficient to describe the low frequency reflectivity, in agreement with \citeauthor{Mardele.2020} \cite{Mardele.2020}.
The emergence of the additional phonon at higher pressures is captured by an additional Lorentzian oscillator in these fits.
For both NbIrTe$_4$ [Fig.~\ref{fig:Nbparameters}(c)] and TaIrTe$_4$ [Fig.~\ref{fig:Taparameters}(c)], the plasma frequency $\omega_{p}$ exhibits a sharp decrease above $P_\text{c}$. To clarify the origin of this behavior, we consider the definition of the plasma frequency, $\omega_\text{p}=\sqrt{\frac{n_\text{e} e^2}{\epsilon_0 m^*_\text{e}}}$~\cite{Fox.2010}, which depends on the charge carrier density $n_\text{e}$ and the effective mass $m^*_\text{e}$. This relationship implies that the reduction in $\omega_\text{p}$ originates from either a decrease in $n_\text{e}$ or an increase in $m^*_\text{e}$. The latter would furthermore cause a decrease in carrier mobility $\mu_{\mathrm{e}}$, as discussed in Ref.~\cite{Koepf.2024}.
As presented in Fig.~\ref{fig:Nbparameters}(d) for NbIrTe$_4$ and Fig.~\ref{fig:Taparameters}(d) for TaIrTe$_4$, the center frequencies of the Lorentzian contributions L1 and L3 (L2 is excluded from further analysis, since it bridges the interpolated region of the multiphonon absorption of the diamond) exhibit a notable change in their pressure dependence at $P_\text{c}$, visible as a distinct alteration in slope.
The spectral weight transfer from low to high frequencies (at around 300\,\cm) occurring above $P_\text{c}$ in both compounds is accompanied by a sharp decrease in the Drude contribution and a significant increase in the interband optical conductivity $\sigma_{1,\text{inter}}$ in the frequency range 250 - 2000\,\cm [see Figs.~\ref{fig:Nbparameters}(e) and \ref{fig:Taparameters}(e)].
This behavior signals a pressure-induced phase transition in both compounds, with a common underlying mechanism driving these changes.

To decide whether the observed changes in the optical conductivity are related to a structural or an electronic phase transition induced by external pressure, we carried out pressure-dependent Raman scattering measurements at room temperature for both materials under the very same experimental conditions (same DAC filling for infrared and Raman measurements). The results are shown in Fig.~S\ref{fig:Ramanspectra} in the Supplemental Material \cite{suppl} (see also Refs.\ \cite{Zhang.2022,Shojaei.2021,Bainsla.2024,Zhang.2023,Chen.2025,Togo.2015} therein). For NbIrTe$_4$ and TaIrTe$_4$, changes in Raman mode positions and mode splittings occur in the critical pressure region $P_c\sim7\text{--}8$\,GPa.
However, we do not observe clear signatures expected for a first-order structural phase transition, such as pronounced discontinuities or the appearance/disappearance of multiple modes. Hence, our data are consistent with a predominantly electronic transition. However, a subtle second-order phase transition, such as structural distortion, cannot be ruled out.

\subsection{Theoretical results}
For the interpretation of our optical conductivity data under pressure, we carried out density functional theory (DFT) calculations to determine the evolution of the electronic structure.
As the two materials produce very similar results, the following section focuses on NbIrTe$_4$. The same graphs can be found in the Supplemental Material for TaIrTe$_4$ (Figs. S\ref{fig:TaBandstructureFermisurface} and S\ref{fig:TaDOS}).

\begin{figure*}[t]
\includegraphics[width=\textwidth]{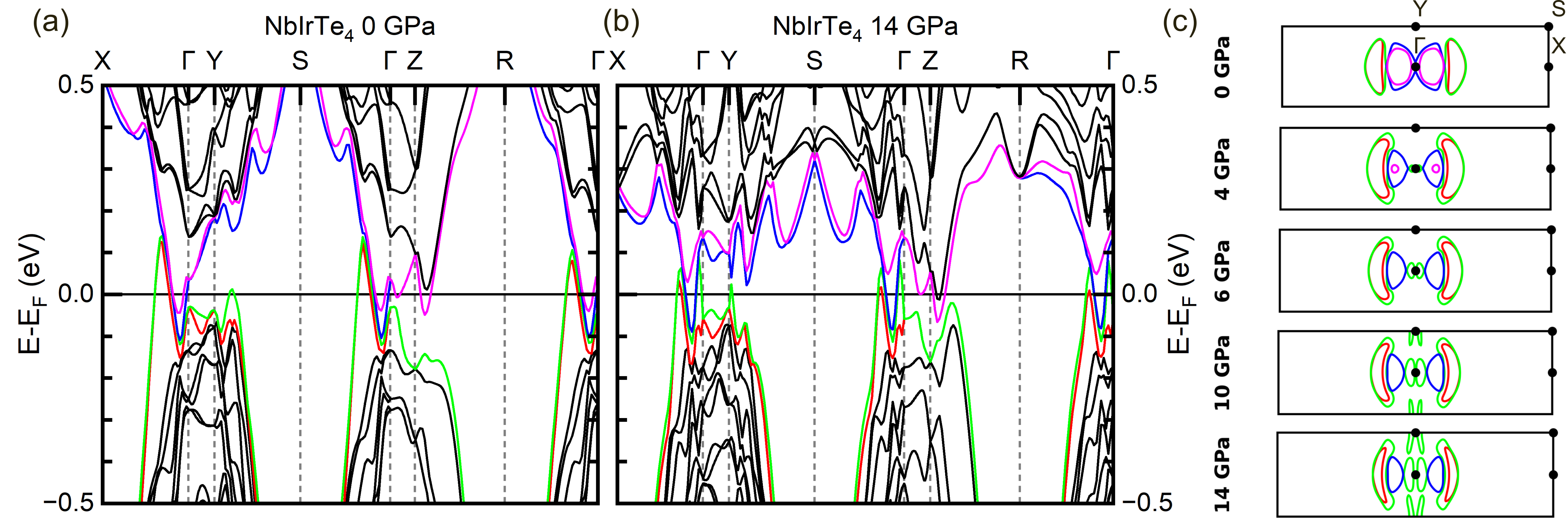}
\caption{Calculated band structure of NbIrTe$_4$ at ambient pressure (a) and at 14$\,$GPa (b). (c) Evolution of the Fermi surface cross-sections under hydrostatic pressure. The panels display the Fermi contours within the Brillouin zone at pressures of 0, 4, 6, 10, and 14$\,$GPa. The high-symmetry points are labeled as $\Gamma$, X, S, and Y. Different colors represent distinct electronic bands crossing the Fermi level (Green and red are electron bands, while blue and pink are hole bands).}\label{fig:NbBandstructureFermisurface}
\end{figure*}

\begin{figure*}[t]
\includegraphics[width=0.8\textwidth]{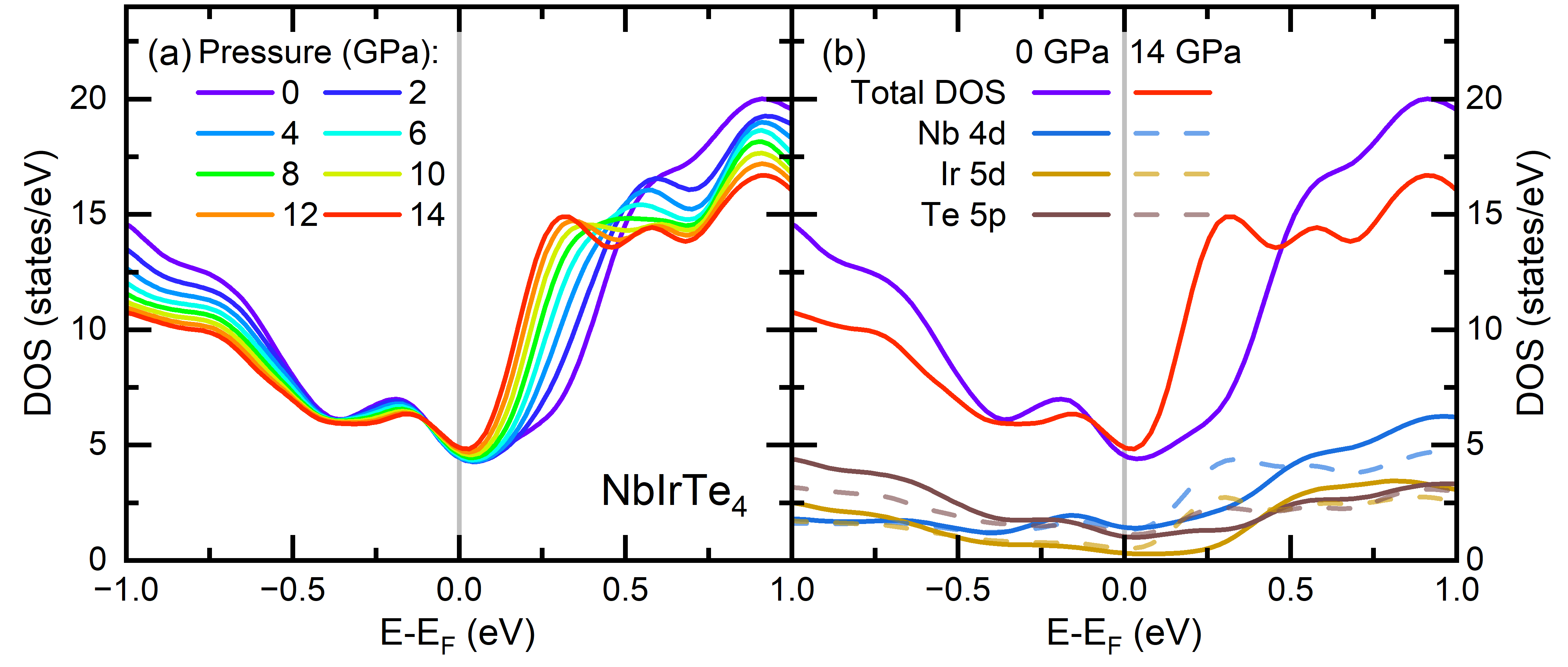}
\caption{Pressure evolution of the electronic density of states (DOS) of $\text{NbIrTe}_4$. (a) Total DOS calculated for hydrostatic pressures ranging from 0 to 14\,GPa. (b) Projected density of states (PDOS) resolved into the main orbital contributions (Nb $4d$, Ir $5d$, and Te $5p$). Solid lines represent the ambient pressure state (0\,GPa), while dashed lines correspond to the high-pressure state (14\,GPa).}\label{fig:NbDOS}
\end{figure*}

The pressure-induced evolution of the electronic band structure for NbIrTe$_4$ is shown in Fig.~\ref{fig:NbBandstructureFermisurface}. Comparing the electronic structure at 0\,GPa [Fig.~\ref{fig:NbBandstructureFermisurface}(a)] and 14\,GPa [Fig.~\ref{fig:NbBandstructureFermisurface}(b)] reveals observable energy shifts of the bands in the vicinity of the Fermi level induced by pressure. These changes manifest clearly in the Fermi surface topology displayed in Fig.~\ref{fig:NbBandstructureFermisurface}(c). As pressure increases, the central crescent-shaped hole bands (blue and pink contours) decrease in size (blue), while the pink contour vanishes from this cross-sectional plane. Simultaneously, the electron bands (red and green contours) decrease, with the green pockets separating into smaller, distinct segments.
This sequence of qualitative changes of the Fermi-surface sheets hints at the possibility of multiple Lifshitz transitions upon compression.

To understand the impact of these changes on the available electronic states, we calculated the density of states (DOS), shown in Fig.~\ref{fig:NbDOS}. Interestingly, the total DOS at the Fermi level ($E_\text{F}$) remains remarkably constant at approximately 5\,states/eV across the entire pressure range [Fig.~\ref{fig:NbDOS}(a)]. However, the profile of the DOS in the conduction band changes significantly. As pressure increases to 14\,GPa, the DOS exhibits a sharp accumulation of states in the narrow energy window of $0.2 - 0.4$\,eV, while the DOS density immediately above this range ($>0.5$\,eV) decreases. The projected DOS analysis in Fig.~\ref{fig:NbDOS}(b) identifies that these changes are driven primarily by the Nb $4d$ and Ir $5d$ orbitals.

\begin{figure*}[t]
\includegraphics[width=0.9\textwidth]{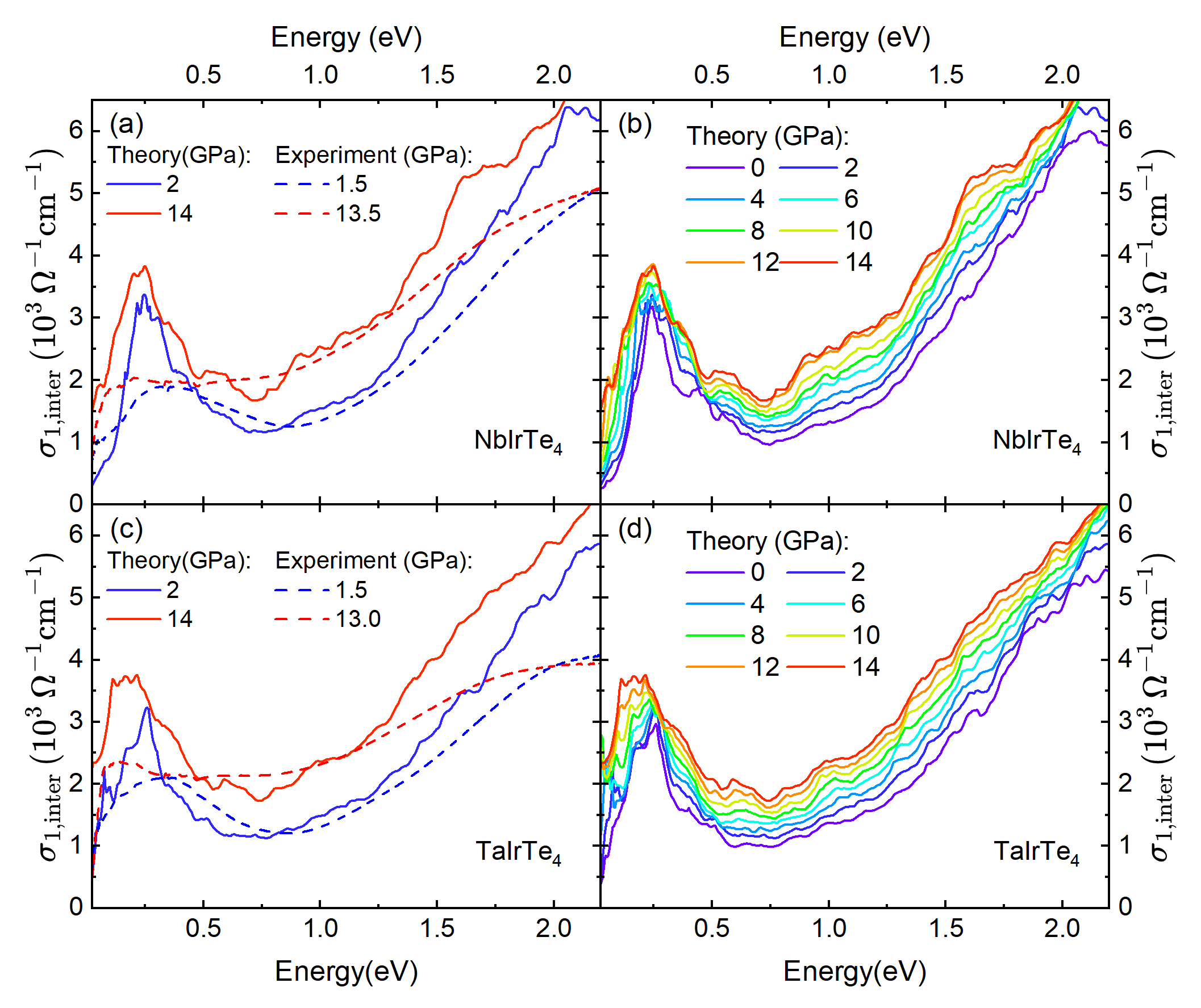}
\caption{(a),(c) Comparison of interband optical conductivity $\sigma_{1,\text{inter}}$ of NbIrTe$_4$ and TaIrTe$_4$, resp., measured at lowest and highest pressure, with the corresponding DFT calculated $\sigma_{1,\text{inter}}$ spectra. (b),(d) Theoretical 
$\sigma_{1,\text{inter}}$ spectrum of NbIrTe$_4$ and TaIrTe$_4$, resp., as a function of pressure.}\label{fig:interbandTheory}
\end{figure*}

Based on this electronic structure, we calculated the interband optical conductivity, which is depicted in Figs.~\ref{fig:interbandTheory}(b) for NbIrTe$_4$ and (d) for TaIrTe$_4$. We include a comparison between the theoretical spectra and the experimental interband conductivity $\sigma_{1,\text{inter}}$ (obtained by subtracting the Drude component) in Figs.~\ref{fig:interbandTheory}(a) and (c). The theoretical $\sigma_{1,\text{inter}}$ spectra show good agreement with the experimental results and reflect the important pressure-induced changes:
approximately no increase of $\sigma_{1,\text{inter}}$ at around 0.4\,eV for NbIrTe$_4$ (at 0.3\,eV for TaIrTe$_4$) and an increase of $\sigma_{1,\text{inter}}$ below and above this point.

These calculations allow us to disentangle the origins of the observed optical features.
The experimentally observed reduction in the Drude term cannot be attributed to a decrease in the electronic DOS at $E_\text{F}$, as this value remains roughly constant in our calculations.
Instead, it is linked to the topological changes of the Fermi surface [Fig.~\ref{fig:NbBandstructureFermisurface}(c)], where the fragmentation and shrinking of the pockets imply a reduction in the Fermi velocity or scattering phase space.
Regarding the interband transitions, the behavior is governed by the energy-dependent changes of the DOS. The emergence of the Nb $4d$ and Ir $5d$ peak at $\sim$~0.3\,eV provides a high density of final states for low-energy excitations, explaining the enhanced conductivity observed in the corresponding infrared region.

\section{DISCUSSION}
Our spectroscopic investigation identifies a critical pressure region $P_c\sim7\text{--}8$\,GPa for both NbIrTe$_4$ and TaIrTe$_4$, characterized by a clear redistribution of spectral weight and a sharp reduction in free carrier response.
To address its origin, we performed high-pressure Raman scattering measurements under identical experimental conditions (see Supplemental Material). These measurements do not show evidence for a first-order structural phase transition. This observation aligns well with previous high-pressure Raman studies on NbIrTe$_4$, which reported no major structural changes up to 40.1\,GPa~\cite{Mu.2021}.
Nevertheless, a more subtle structural change, such as a slight distortion, cannot be ruled out.
In particular, we do not observe the pronounced, non-continuous Raman changes associated with the pressure-induced $T_d\rightarrow$~1T$'$ structural transition reported for the related compound WTe$_2$~\cite{Zhou.2016}, supporting a predominantly electronic origin of the transition in NbIrTe$_4$/TaIrTe$_4$.
Similar anomalies in the pressure dependence of Raman modes (e.g., changes of pressure coefficients/slopes and linewidth anomalies), discussed as fingerprints of pressure-induced electronic (Lifshitz-type) transitions without a major structural transformation, were reported for the Weyl semimetals NbP and NbAs~\cite{Gupta.2018NbP, Gupta.2018NbAsTaAs} and for 2H-MoTe$_2$~\cite{Bera.2017MoTe2}.
For comparison, in the related compound WTe$_2$ a pressure-induced structural phase transition from the orthorhombic $T_d$ phase to the monoclinic 1T$'$ phase (space group $P2_1/m$) was identified by synchrotron x-ray diffraction and Raman spectroscopy, accompanied by a $\sim$20.5\% collapse of the unit-cell volume~\cite{Zhou.2016}.
In that case, the Raman response exhibits pronounced mode splitting and non-continuous changes across the transition, unlike what we observe here for NbIrTe$_4$ and TaIrTe$_4$. The absence of a comparably pronounced structural anomaly in available high-pressure x-ray data for NbIrTe$_4$ supports our interpretation of a predominantly electronic transition at $P_\text{c}$~\cite{Mu.2021}.
Consequently, the anomalies observed in our optical data at $P_\text{c}$ are most likely attributed to an electronic phase transition, involving a significant reconstruction of the Fermi surface.
This interpretation is strongly supported by high-pressure electrical transport measurements, which reveal a complex response to compression in this material class. In NbIrTe$_4$, Hall effect measurements indicate a transition from a multiband character to a hole-dominated state between 4.1 and 12\,GPa~\cite{Jin.2021, Mu.2021}.
Ref.~\cite{Mu.2021} reports a nonmonotonic pressure dependence of the resistivity with a minimum at $\sim$12\,GPa, coincident with a maximum in the hole concentration. They suggest that this phenomenon could potentially indicate a topological phase transition, pressure-induced multiband competition or a pressure-induced Lifshitz transition.
More generally, in the pressure range relevant to the present measurements, the published room-temperature transport data show a rather heterogeneous behavior. For NbIrTe$_4$, ranging from a nonmonotonic resistivity trend~\cite{Mu.2021} to a monotonic decrease of the measured resistance~\cite{Long.2021}, whereas the available transport data for TaIrTe$_4$ also show a continuous decrease of the measured resistance under pressure in the corresponding low-pressure range~\cite{Cai.2019}.
Furthermore, we observe a decrease in the plasma frequency $\omega_\text{p}$ above $P_\text{c}$, signaling a reduction in metallic character. 
This aligns with the non-monotonic evolution of charge carrier density reported in Hall studies, where an initial decrease is followed by a regime change to hole-dominated transport~\cite{Mu.2021,Jin.2021}.
Although our DFT calculations suggest a roughly constant total density of states at $E_\text{F}$, the reduction in $\omega_\text{p}$ can be understood as a consequence of topological changes affecting the carrier mobility or scattering phase space, specifically, the fragmentation and shrinking of Fermi pockets (see Fig.~\ref{fig:NbBandstructureFermisurface}).
Consistently, the DFT-calculated Raman-active phonon modes of NbIrTe$_4$ (Supplemental Material, Fig.~S\ref{fig:DFTPhononModes}) exhibit mode merging in the same critical pressure range as the measured Raman modes, which is coincident with the pressure range where we infer a reconstruction of the Fermi surface. Taken together, these observations suggest that both the Fermi-surface reconstruction and the phonon-mode merging are closely linked to the transition at $P_\text{c}$ and may be connected to the changes of the Drude response.
Such purely electronic transitions involving topological changes to the Fermi surface are characteristic of Lifshitz transitions~\cite{Lifshitz.1960}, a phenomenon recently identified under pressure in the related layered Dirac semimetal ZrSiTe~\cite{Krottmüller.2020, Ebad-Allah.2019}.
In the broader context, pressure serves as a potent tuning parameter for the topological landscape of these materials.

NbIrTe$_4$ is predicted to host 16 Weyl points (WPs) due to weak spin-orbit coupling~\cite{Li.2017, Zhou.2019, Schoenemann.2019}. Theoretical calculations suggest these WPs are remarkably robust, persisting even under significant volume reduction, while the Fermi surface evolves toward the hole-dominated state observed experimentally~\cite{Mu.2021}.
Similarly, TaIrTe$_4$, which hosts 12 WPs and robust nodal lines~\cite{Koepernik.2016, Zhou.2018}, is predicted to undergo significant expansion of its nodal loops under compression.

While the low-energy optical response of TaIrTe$_4$ at 1.5\,GPa reproduces the linear-in-frequency dependence observed at ambient pressure, the interpretation of this feature requires careful consideration.~\citeauthor{Mardele.2020} modeled this linearity ($<$40 meV) as a signature of interband transitions within tilted Weyl cones. However, they explicitly noted that such a response is ``not a smoking gun'' for Weyl dispersion, as an energy-dispersive nodal line at the Fermi level would yield a similar quasi-linear optical conductivity followed by a high-energy plateau.
Our high-pressure data are consistent with this alternative scenario, which highlights the plateau behaviour above 40\,meV. This range was not described in detail in the ambient pressure study.
The emergence of this flat, plateau-like response, which extends up to the diamond absorption edge in TaIrTe$_4$, aligns strongly with the theoretical predictions of dispersive nodal lines coexisting with Weyl points in this material family~\cite{Zhou.2018, Li.2017, Zhou.2019}. Furthermore, the evolution of this feature under compression offers physical insight into the topological structure: the increase in the plateau level suggests an elongation of the nodal lines, consistent with the predicted expansion of nodal loops in TaIrTe$_4$ under hydrostatic pressure~\cite{Zhou.2018}.

Finally, we would like to comment on the interlayer interaction as the potential driving force of the phase transition in TaIrTe$_4$ and NbIrTe$_4$. According to Ref.\ \cite{Cai.2019}, the lattice parameter ratio c/a of TaIrTe$_4$ first decreases modestly under pressure up to $\sim$7~GPa, and then drastically decreases above 7~GPa, implying a strong increase in interlayer interaction above this critical pressure \cite{Cai.2019}.
Thus, we relate the pressure-induced phase transition at P$_c$=7-8\,GPa to the enhanced interlayer interaction.
Ultimately, our results demonstrate that hydrostatic pressure effectively tunes the van der Waals interlayer coupling~\cite{Liu.2017}, as illustrated in Fig.~\ref{fig:crystalstructure}(b), driving these systems through a Lifshitz-type transition and suppressing the ambient-pressure magnetoresistive state, often as a precursor to the superconductivity reported at higher pressures~\cite{Cai.2019, Long.2021}.

\section{CONCLUSION}
In conclusion, our high-pressure spectroscopic investigation of NbIrTe$_4$ and TaIrTe$_4$ single crystals provides compelling evidence for a pressure-induced, reversible, phase transition, most likely electronic in nature, in both van der Waals semimetals. Infrared spectroscopy supplemented by Raman scattering measurements reveal anomalies at the common critical pressure $P_\text{c}=7\text{--}8$\,GPa.
The Raman spectra show no evidence of a first-order structural transition (or abrupt symmetry change), though more subtle effects (e.g. second-order) cannot be excluded. In particular, we do not observe the pronounced, non-continuous Raman changes associated with the pressure-induced T$_d$--1T$'$ structural transition reported for the related compound WTe$_2$~\cite{Zhou.2016}, supporting a predominantly electronic origin of the transition in NbIrTe$_4$/TaIrTe$_4$. The phase transition is characterized by a significant redistribution of optical spectral weight, a sharp reduction in the free carrier concentration, and the emergence of a phonon mode, which was screened by free charge carriers at pressures below $P_\text{c}$. These findings signal a pressure-induced reduction of the metallic character at $P_\text{c}$.

Our ab initio calculations affirm these experimental observations, reproducing the main pressure-dependent trends of the optical conductivity. The theoretical analysis suggests that the observed reduction in Drude weight is not primarily driven by a loss of electronic states at the Fermi level, which remains nearly constant, but rather by significant topological modifications of the Fermi surface, specifically the shrinking and fragmentation of the electron and hole pockets. Furthermore, the calculations identify that the enhanced interband conductivity arises from a pressure-induced accumulation of Nb $4d$ and Ir $5d$ states in the conduction band, which increases the joint density of states available for low-energy optical transitions.

Our work establishes pressure as a key tool to tune the electronic and topological properties of these Weyl semimetals via a Lifshitz-type transition.

\begin{acknowledgments}
M.L.\ acknowledges technical support by Beate Sp\"orhase.
C.K.~acknowledges financial support by the Deutsche Forschungsgemeinschaft (DFG), Germany, through Grants No.~KU 1432/15-1.
C.K. and L.C.\ acknowledge support by the DFG under Grant No. TRR 360 – 492547816 (subprojects A1 and A5).
L.B.~acknowledges support from the US DoE, BES program, through award DE-SC0002613. The National High Magnetic Field Laboratory acknowledges support from the US-NSF Cooperative agreement Grant DMR-2128556, and the state of Florida.
The TaIrTe$_4$ single crystal growth work was performed at the Pennsylvania State University Two-Dimensional Crystal Consortium–Materials Innovation Platform (2DCC-MIP), which is supported by NSF Cooperative Agreement No. DMR-2039351
\end{acknowledgments}

\end{document}